 \documentclass[final,3p,times]{elsarticle}
\usepackage{graphicx}
\usepackage{amssymb}
\usepackage{tabularx}
\usepackage{newtxmath}
\usepackage{mathbbol}

\usepackage{lineno}

\usepackage{hyperref}

\newcommand{\beq}{\begin{equation}}
\newcommand{\eeq}{\end{equation}}

\usepackage{comment}
\usepackage{amsmath}
\usepackage{lipsum}
\usepackage{graphics}
\usepackage[dvipsnames]{xcolor}
\usepackage{ulem}

\journal{arXiv}

\begin{document}

\begin{frontmatter}

%% Title, authors and addresses

\title{Data-driven Modeling of the Mechanical Behavior of Anisotropic Soft Biological Tissue}

%% use the tnoteref command within \title for footnotes;
%% use the tnotetext command for the associated footnote;
%% use the fnref command within \author or \address for footnotes;
%% use the fntext command for the associated footnote;
%% use the corref command within \author for corresponding author footnotes;
%% use the cortext command for the associated footnote;
%% use the ead command for the email address,
%% and the form \ead[url] for the home page:
%%
%% \title{Title\tnoteref{label1}}
%% \tnotetext[label1]{}
%% \author{Name\corref{cor1}\fnref{label2}}
%% \ead{email address}
%% \ead[url]{home page}
%% \fntext[label2]{}
%% \cortext[cor1]{}
%% \address{Address\fnref{label3}}
%% \fntext[label3]{}

%% use optional labels to link authors explicitly to addresses:
%% \author[label1,label2]{<author name>}
%% \address[label1]{<address>}
%% \address[label2]{<address>}

\author{Vahidullah Tac$^{1}$, Vivek D. Sree$^{1}$, Manuel K. Rausch$^{2}$, and Adrian B. Tepole$^{1,3}$ }

\address{$^1$School of Mechanical Engineering, Purdue University, West Lafayette, IN, USA\\ $^2$Department of Aerospace Engineering and Engineering Mechanics, the University of Texas at Austin, Austin, TX, USA\\$^3$Weldon School of Biomedical Engineering, Purdue University, West Lafayette, IN, USA}

\begin{abstract}
Constitutive models that describe the mechanical behavior of soft tissues have advanced greatly over the past few decades. These expert models are generalizable and require the calibration of a number of parameters to fit experimental data. However, inherent pitfalls stemming from the restriction to a specific functional form include  poor fits to the data, non-uniqueness of fit, and high sensitivity to parameters. In this study we design and train fully connected neural networks as material models to replace or augment expert models. To guarantee objectivity, the neural network takes isochoric strain invariants as inputs, and outputs the value of strain energy and its derivatives with respect to the invariants. Convexity of the material model is enforced through the loss function. Direct prediction of the derivative functions -rather than just predicting the energy- serves two purposes: it provides flexibility during training, and it enables the calculation of the elasticity tensor through back-propagation.  We showcase the ability of the neural network to learn the mechanical behavior of porcine and murine skin from biaxial test data. Crucially, we show that a multi-fidelity scheme which combines high fidelity experimental data with low fidelity analytical data yields the best performance. The neural network material model can then be interpreted as the best extension of an expert model: it learns the features that an expert has encoded in the analytical model while fitting the experimental data better. Finally, we implemented a general user material subroutine (UMAT) for the finite element software Abaqus and thereby make our advances available to the broader computational community. We expect that the methods and software generated in this work will broaden the use of data-driven constitutive models in biomedical applications.  
\end{abstract}


\end{frontmatter}

%%%%%-------- SECTION ---------%%%%
\section*{Introduction}\label{motiv}
%%%%%--------------------------%%%%

%% Skin is important, mechanical behavior is characterized with analytical functions but there is no best one.  
%% (why is this important, what we know, what is the gap) 
Skin is the largest organ in the body and understanding its mechanical properties is a crucial step in many biomedical applications, from the design of prostheses to surgical intervention \cite{buganza15, neumann_expansion}. The tissue microstructure is characterized by the presence of semi-flexible biopolymer fiber networks such as collagen and elastin, which endow skin with nonlinear and anisotropic mechanical behavior \cite{sherman2017}. The mechanical properties of skin are actually common across many collagenous soft connective tissue \cite{kakaletsis2021,meador2020b}. Traditionally, the mechanical behavior of skin and other soft tissues has been modelled using expert-constructed constitutive equations \cite{holzapfel2000, holzapfel2009, gasser2005GOH, Humphrey1990}. In this approach, a functional form describing the main features of the mechanical behavior of a family of materials is constructed first. Then, the free parameters in the equations are fitted to a specific material in the family to obtain a calibrated model. Thus, this approach necessitates choosing a specific functional form for the constitutive model. In turn, inherent restrictions of the functional form can result in poor fitting and poor predictions. Unfortunately, even restricting our attention to skin out of all connective soft tissue, there is currently no consensus on the choice of model that is most suitable in a particular application \cite{shergold_pigskin, fung_biomechanics, delalleau_comparison, Lapeer_reduced_poly, Joodaki_review, meador2020, lee2020, garikipati2011perspectives, lee2018}. 
%For example, models originally developed for rubbers can be fitted reasonably well for specific loading conditions, exponential functions and one or two representative fiber families perform well in a wide variety of deformation regimes with few parameters [cite], models that rely on probability density distributions for orientation and undulation of collagen fibers are also suitable [cite].

%%  data-driven approaches new ways of modeling materials without prescribing analytical form, so more freedom 
A new, emergent approach to material modeling is the use of data-driven methods \cite{garikipati2011perspectives, Rahmani2021hybrid, garikipati2020multiresolution, peng2020multiscale}. Among them, neural network have been used successfully to describe the mechanical behavior of several materials \cite{garikipati2020multiresolution, Vlassis202elastoplast, lu2020multifidelity, lejeune2021transferlearn}. In this approach, the burden of choosing a suitable analytical representation is no longer present, and since the training of the neural network is not restricted to a parametric family of functions, it often results in more accurate predictions than traditional models \cite{leng2021, liu2020, reimann2019}. In particular, a common application of machine learning has been to create macroscale constitutive models out of micromechanical simulations \cite{reimann2019, lejeune2021transferlearn}. 

One potential drawback of this approach is scarcity of high fidelity test data to train the data-driven models. It is often difficult, expensive, and sometimes impossible to gather large amounts of data to train a neural network over the entire input space of deformations. One approach to overcome this problem is to use neural networks in a similar vein as the traditional model development, i.e., training a neural network using data from multiple specimens and multiple loading conditions in order to learn an average response. Then, data of a single material can be fitted by adjusting only a few parameters of the neural network  \cite{liu2020}. 
%This approach still requires large amounts of training data, which may not be available in every case. Additionally, this method still involves processing two separate sets of data consecutively and two separate optimization processes. 

The route we follow here lies between the purely data-driven approach and the expert modeling approach. Expert models already include knowledge of physics relevant to soft tissue, observations of both the mechanical behavior and underlying microstructure properties, and intuition from the modeller regarding the main features of the material response. For example, to model skin, we have restricted our attention to the framework of hyperelasticity and used expert-designed strain energy functions to fit murine and porcine skin data \cite{meador2020, tinareview}. However, the error in the fits can be undesirable, the parameters non-unique, and the predictions can be highly sensitive to the material parameters, especially outside of the loading regime used during calibration. Here we propose to use the analytical strain energy functions as low fidelity approximations, and complement the data with high fidelity experimental measurements. This approach is based on the recent literature that shows the advantage of multi-fidelity schemes over single fidelity approaches \cite{lejeune2021transferlearn, lu2020multifidelity, perdikaris2020multifidelity, lee2020}.  

Specifically, we train fully connected neural networks that output the strain energy and its derivatives with respect to the isochoric strain invariants, including the effect of anisotropy. The training of the neural network is done with a multi-fidelity dataset that includes low fidelity data from conventional material models, and high fidelity data from biaxial experiments.  Furthermore, the loss function is designed to impose convexity constraints on the strain energy, which ensures physically meaningful elasticity tensors. Finally, we implemented a user material (UMAT) subroutine  for the widely used finite element package Abaqus \cite{abaqus}. The UMAT reads all the parameters of a fully connected neural network from the input file and evaluates the neural network to construct the stress and elasticity tensors  needed in the finite element simulation. The work shown here will extend the reach of machine learning tools to improve the modeling of soft tissue mechanics, in particular through improved constitutive models ready to be used in commercial finite element codes.

%%------------------------------------------------------%%
% Fig 1: Diagram
\begin{figure}[h!]
\centering
\includegraphics[width=0.9\linewidth]{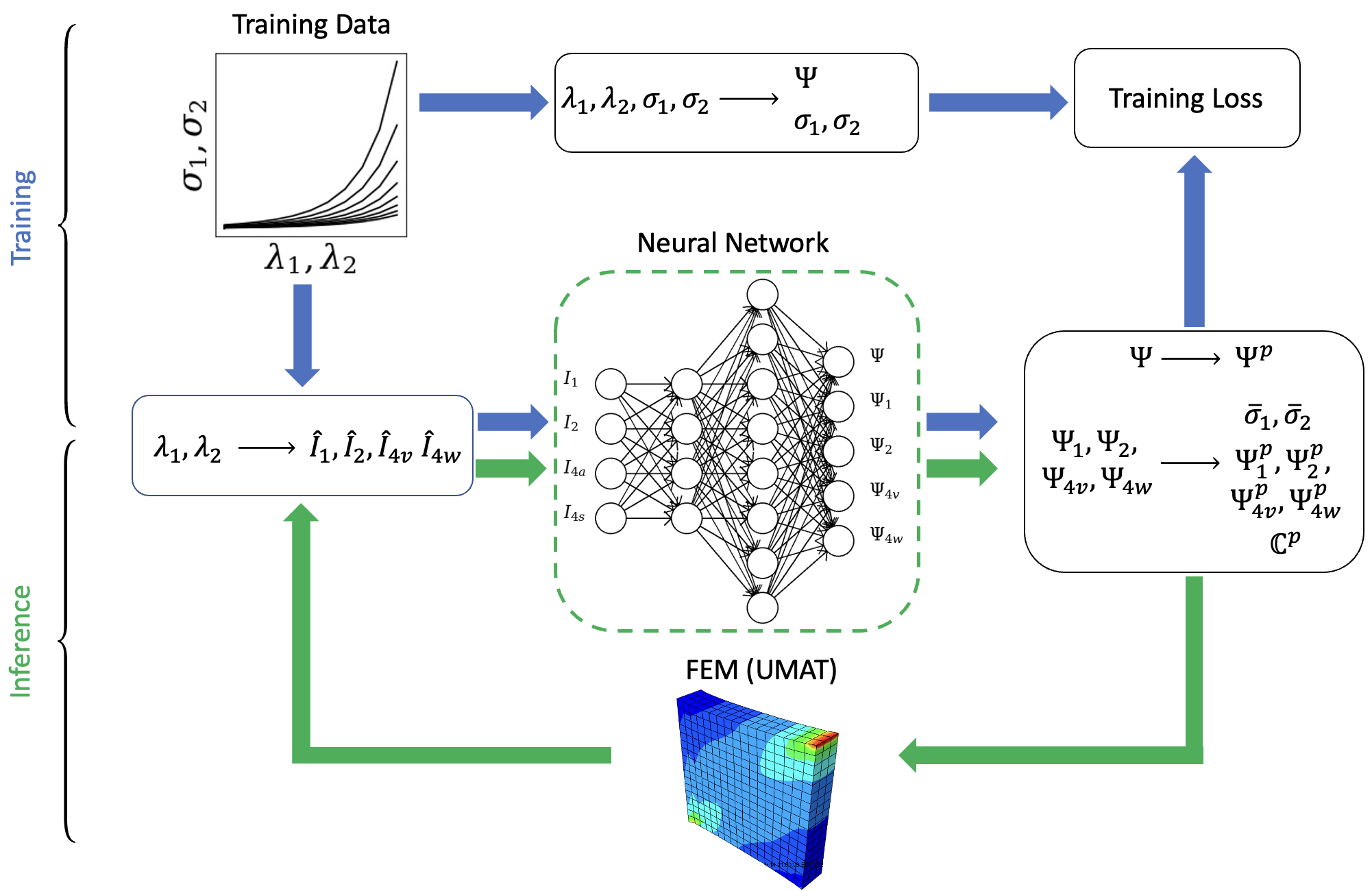}
\caption{Diagram depicting the training and inference processes of the neural network material model.}
\label{fig01} 
\end{figure}
%%------------------------------------------------------%%

%%%%%-------- SECTION ---------%%%%
\section*{Materials and Methods}
%%%%%--------------------------%%%%

%%%%%----- SUBSECTION ---------%%%%
\subsection*{Constitutive equations for a hyperelastic material with two families of fibers} 
%%%%%--------------------------%%%%
In this study we use a form of Helmholtz free energy, $\Psi$, that is a function of the right Cauchy-Green deformation tensor, $\mathbf{C}$, and two material direction vectors in the reference configuration, $\mathbf{v}_0$ and $\mathbf{w}_0$. This form of the Helmholtz free energy function allows for greater flexibility in recreating the mechanical behavior of materials where more than one family of fibers is present or even when the orientation of fibers is random, which is usually the case in biological tissues. For soft tissues, which we assume to be nearly incompressible, the additive split into isochoric and volumetric parts is used \cite{holzapfel2000},
\begin{equation}
\Psi = \Psi_{\text{iso}}(\hat{I}_1, \hat{I}_2, \hat{I}_{4v}, \hat{I}_{4w}) + \Psi_{\text{vol}}(J)\, ,
\end{equation}
where $J=\sqrt{\det{\mathbf{C}}}$ is the volume change, and the isochoric strain invariants, $\hat{I}_1$, $\hat{I}_2$, $\hat{I}_{4v}$ and $\hat{I}_{4w}$ are defined as 

\begin{equation}
    \begin{aligned}
        \hat{I}_1 &= \hat{\mathbf{C}}:\mathbf{I} = tr(\hat{\mathbf{C}}) ,\\
        \hat{I}_2 &= \frac{1}{2}[\hat{I}_1^2 - tr(\hat{\mathbf{C}}^2)] ,\\
        \hat{I}_{4v} &= \hat{\mathbf{C}}:\mathbf{v}_0\otimes\mathbf{v}_0 = \hat{\mathbf{C}}:\mathbf{V}_0 ,\\
        \hat{I}_{4w} &= \hat{\mathbf{C}}:\mathbf{w}_0\otimes\mathbf{w}_0 = \hat{\mathbf{C}}:\mathbf{W}_0 \, .
    \end{aligned}
    \label{def_iso_invs}
\end{equation}
The isochoric right Cauchy Green deformation, $\hat{\mathbf{C}}$, can be defined in terms of the isochoric part of the deformation gradient,
\begin{align}
    \hat{\mathbf{F}} &= J^{-1/3}\mathbf{F}  ,\\
    \hat{\mathbf{C}} &= \hat{\mathbf{F}}^\top\hat{\mathbf{F}}  = J^{-2/3}\mathbf{C} \, .
    \label{def_iso_C}
\end{align}
The second Piola-Kirchhoff stress tensor, $\mathbf{S}$, follows from the Doyle-Erickson formula by differentiating the strain energy $\Psi$ with respect to $\mathbf{C}$ and following the procedure outlined by Coleman and Noll \cite{DOYLE195653, Coleman_Noll1974}, arriving at
\begin{equation}
    \mathbf{S} = 2\frac{\partial\Psi}{\partial \mathbf{C}} = \mathbf{S}_{\text{iso}} + \mathbf{S}_{\text{vol}} \, ,
    \label{def_S}
\end{equation}
\begin{equation}
    \mathbf{S}_{\text{iso}} = \hat{\mathbf{S}}:\frac{\partial \hat{\mathbf{C}}}{\partial \mathbf{C}} = J^{-2/3} \hat{\mathbf{S}}:\mathbb{P}_1,  \;\;\; \mathbf{S}_{\text{vol}} =  2p\frac{\partial J}{\partial \mathbf{C}} = Jp\mathbf{C}^{-1} \, .
\end{equation}
The following definition of the pressure has been introduced $p = d \Psi_{\text{vol}} / dJ$. Additionally, the fictitious second Piola-Kirchhoff stress tensor, $\hat{\mathbf{S}}$, is the result from differentiating the isochoric part of the strain energy with respect to the isochoric invariants, i.e. $\hat{\Psi}_1 = \partial \Psi_{\text{iso}} / \partial \hat{I}_1$, $\hat{\Psi}_2 = \partial \Psi_{\text{iso}} / \partial \hat{I}_2$, $\hat{\Psi}_{4v} = \partial \Psi_{\text{iso}} / \partial \hat{I}_{4v}$ and $\hat{\Psi}_{4w} = \partial \Psi_{\text{iso}} / \partial \hat{I}_{4w}$. The full expansion of the fictitious stress tensor is

\begin{equation}
    \hat{\mathbf{S}} = 2\frac{\partial \Psi_{\text{iso}}}{\partial \hat{\mathbf{C}}}  = 2[\hat{\Psi}_1 \mathbf{I} + \hat{\Psi}_2 (\hat{I}_1 \mathbf{I} - \hat{\mathbf{C}}) + \hat{\Psi}_{4v} \mathbf{V}_0 + \hat{\Psi}_{4w} \mathbf{W}_0 ] \, .
    \label{def_Shat}
\end{equation}
The term \textit{fictitious} originates from the fact that derivatives with respect to the full right Cauchy Green deformation tensor needs the projection with the fourth order tensor 
\begin{equation*}
    \mathbb{P}_1 = \mathbb{I} - \frac{1}{3}\mathbf{C}^{-1} \otimes \mathbf{C} \, ,
\end{equation*}
which relates derivatives with respect to the isochoric part of the deformation to derivatives with respect to the total deformation. 
If the material under consideration is exactly incompressible, i.e. $J=1$, then $\hat{\mathbf{C}} = \mathbf{C}$, (\ref{def_S}) reduces to 
\begin{equation}
    \mathbf{S} = 2[\hat{\Psi}_1 \mathbf{I} + \hat{\Psi}_2 (\hat{I}_1 \mathbf{I} - \mathbf{C}) + \hat{\Psi}_{4a} \mathbf{V}_0 + \hat{\Psi}_{4s} \mathbf{W}_0 ] + p\mathbf{C}^{-1} \, ,
    \label{S_incomp}
\end{equation}
and the pressure becomes an unknown Lagrange multiplier field. 

The UMAT subroutine requires the computation of the Cauchy stress tensor, $\mathbf{\sigma}$. Thus, for completeness, we state the standard push forward operation for the stress
\begin{equation}
    \mathbf{\sigma} = \frac{1}{J} \mathbf{F}\mathbf{S}\mathbf{F}^\top \, .
    \label{S_to_sigma}
\end{equation}

The finite element subroutine also requires the computation of the elasticity tensor, $\mathbb{c}_{\text{abaqus}}$ \cite{nele2020}. For ease of derivation, the material version of the elasticity tensor, $\mathbb{C}$, is introduced first,

\begin{equation}
    \mathbb{C} = 2\frac{\partial \mathbf{S}}{\partial \mathbf{C}} = \mathbb{C}_{\text{iso}} + \mathbb{C}_{\text{vol}} \, ,
    \label{def_CC}
\end{equation}
\begin{equation}
   \mathbb{C}_{\text{iso}} = 2\frac{\partial \mathbf{S}_{\text{iso}}}{\partial \mathbf{C}}, \;\;\; \mathbb{C}_{\text{vol}} = 2\frac{\partial \mathbf{S}_{\text{vol}}}{\partial \mathbf{C}} \, . 
\end{equation}

The expressions for the volumetric and isochoric parts of the elasticity tensor, $\mathbb{C}_{\text{vol}}$ and $\mathbb{C}_{\text{iso}}$, can be further expanded,
\begin{equation}
    \mathbb{C}_{\text{vol}} = J\Tilde{p} \mathbf{C}^{-1} \otimes \mathbf{C}^{-1} - 2Jp \mathbf{C}^{-1} \odot \mathbf{C}^{-1} \, , 
\end{equation}
\begin{equation}
    \mathbb{C}_{\text{iso}} = -\frac{2}{3}\mathbf{S}_{\text{iso}} \otimes \mathbf{C}^{-1} + J^{-4/3}\mathbb{P}_1:\hat{\mathbb{C}}:\mathbb{P}_1^T - \frac{2}{3} \mathbf{C}^{-1} \otimes \mathbf{S}_{\text{iso}} + \frac{2}{3} J^{-2/3} tr(\hat{\mathbf{S}}) \mathbb{P}_2 \, ,
    \label{def_CC_iso}
\end{equation}
where the modified pressure term, $\Tilde{p} = p+Jdp/dJ$, has been introduced, as well as the special product noted by ($\odot$) defined as $(\bullet \odot \circ)_{ijkl} = [(\bullet) _{ik} (\circ) _{jl} + (\bullet) _{il} (\circ) _{jk}]/2$, and an additional fourth  order projection tensor $\mathbb{P}_2$,
\begin{equation*}
    \mathbb{P}_2 = \mathbf{C}^{-1} \odot \mathbf{C}^{-1} - \frac{1}{3} \mathbf{C}^{-1} \otimes \mathbf{C}^{-1} \, .
\end{equation*}

Finally, the fictitious elasticity tensor, $\hat{\mathbb{C}}$ is obtained from differentiating the fictitious stress tensor with respect to the isochoric part of the deformation tensor,
\begin{equation*} 
    \hat{\mathbb{C}} = 2\frac{\partial \hat{\mathbf{S}}}{\partial \hat{\mathbf{C}}} \, .
\end{equation*}
The full expansion of $\hat{\mathbb{C}}$ is available in the Appendix. The only remark needed in the main text is that the tensor $\hat{\mathbb{C}}$ requires the second derivatives of the strain energy function with respect to the isochoric invariants: $\hat{\Psi}_{11} = \partial^2 \hat{\Psi} / \partial^2 \hat{I}_1$, $\hat{\Psi}_{12} = \partial^2 \hat{\Psi} / \partial \hat{I}_2 \partial \hat{I}_1$, $\hat{\Psi}_{14v} = \partial^2 \hat{\Psi} / \partial \hat{I}_{4v} \partial \hat{I}_1$, etc. This point will become important in the design in the neural network later on. 

As stated above, the elasticity tensor needed in the UMAT subroutine is associated with the deformed configuration. The push-forward operation for the elasticity tensor yields

\begin{equation}
\mathbb{c} = \frac{1}{J}(\mathbf{F} \overline{\otimes} \mathbf{F}): \mathbb{C}:(\mathbf{F} \overline{\otimes} \mathbf{F})^T
\label{eq_cc}
\end{equation}
where we have introduced the modified dyadic product defined as $(\bullet \overline{\otimes} \circ)_{ijkl} = (\bullet) _{ik} (\circ) _{jl}$. The tensor $\mathbb{c}$ is related to the Truesdell stress rate; however,  Abaqus  increments employ the Jaumann stress rate. Therefore, the consistent tangent for Abaqus is not Eq. (\ref{eq_cc}) but rather

\begin{equation}
\mathbb{c}_{\mathrm{abaqus}} = \mathbb{c}+ \frac{1}{2}(\mathbf{\sigma} \overline{\otimes} \mathbf{I} + \mathbf{\sigma} \underline{\otimes} \mathbf{I} + \mathbf{I} \overline{\otimes} \mathbf{\sigma} + \mathbf{I} \underline{\otimes} \mathbf{\sigma}) \, .
\label{eq_ccabaqus}
\end{equation}

with the modified dyadic product $(\bullet \underline{\otimes} \circ)_{ijkl} = (\bullet) _{il} (\circ) _{jk}$.

Lastly, we remark that during the training of the neural network, incompressibility can be satisfied exactly. In this case $p$ can be determined from boundary conditions. However, in the UMAT we use the following function for $p$,
\begin{equation}
    p = K(J-1)\, ,
\end{equation}
with $K$ the bulk modulus.  In this study we set $K = 10$ kPa. 

%%%%%----- SUBSECTION ---------%%%%
\subsection*{Neural network structure and training}
%%%%%--------------------------%%%%
We use a fully connected neural network to learn the mechanical behavior of skin. The neural network takes four inputs, the isochoric strain invariants  in (\ref{def_iso_invs}), and produces five outputs, the strain energy, $\Psi_{\text{iso}}^{\text{p}}$, and its derivatives with respect to the invariants, $\hat{\Psi}_1^{\text{p}}$, $\hat{\Psi}_2^{\text{p}}$, $\hat{\Psi}_{4v}^{\text{p}}$ and $\hat{\Psi}_{4w}^{\text{p}}$. Note that the notation  $(\bullet)^{\text{p}}$  is used to denote the predicted values of the neural network. The network architecture is summarized in Table \ref{table01}. 

Training data for the neural network is in the form of stretch, stress and strain energy values. Therefore, the first component of the loss function is simply the comparison of the predicted strain energy, ${\Psi}_{\text{iso}}^{\text{p}}$, against the observed $\Psi_{\text{iso}}$. Since the derivatives are direct outputs, a regularization term is added to the loss function to enforce that the predicted derivatives coincide with the derivatives of  $\Psi_{\text{iso}}^{\text{p}}$. Using back-propagation, the derivatives of ${\Psi}_{\text{iso}}^{\text{p}}$ with respect to the inputs are denoted as  $\hat{\Psi}_{i}^{\text{diff}}$, with $i=1,2,4v,4w$. Thus, the first component of the loss function is 

\begin{equation}
      \mathcal{L}_1 = \frac{1}{N} \sum_{n=1}^{N} \left[ \left( \left(\Psi_{\text{iso}}^{\text{p}}\right)^{(n)} - \left(\Psi_{\text{iso}}^{\phantom{p}}\right)^{(n)} \right)^2 +\sum_{i=1,2,4v,4w} \left( \left(\hat{\Psi}_i^{\text{diff}}\right)^{(n)} - \left(\hat{\Psi}_i^{\text{p}\phantom{-}}\right)^{(n)}\right)^2 \right] \, ,
\end{equation}
where $(\bullet)^{(n)}$ denotes the $n^{\text{th}}$ training point, out of a total of $N$ training points. 

The second component of the loss results from comparing the stress computed with the neural network outputs against the observed stress $\mathbf{\sigma}$. The stress, defined in Eqs. (\ref{S_incomp}) and (\ref{S_to_sigma}), is computed based on the neural network outputs ${\Psi}_{\text{iso}}^{\text{p}}$, to produce $\mathbf{\sigma}^{\text{p}}$. The loss can then be simply stated as

\begin{equation}
    \mathcal{L}_2 = \frac{1}{N} \sum_{n=1}^{N} \left|\left| \left({\mathbf{\sigma}^{\text{p}}}\right)^{(n)} - \left(\mathbf{\sigma}\right)^{(n)} \right|\right|_F
\end{equation}
where $||\bullet||_F$ denotes the Frobenius norm.

To ensure optimal convergence of the finite element framework, the strain energy must be a (poly)convex function of the deformation \cite{ehret2007polyconvex}. In the case of a strain energy function defined in terms of the deformation invariants, the strain energy needs to be a convex function of its inputs. For a function to be convex, it's Hessian matrix, $\mathbf{H}$, must be positive semi-definite. The Hessian matrix is 
\begin{equation}
    \mathbf{H} = 
    \begin{pmatrix}
    \Psi_{11}^{\text{diff}} & \Psi_{12}^{\text{diff}} & \Psi_{14v}^{\text{diff}} & \Psi_{14w}^{\text{diff}} \\[8pt]
    \Psi_{21}^{\text{diff}} & \Psi_{22^{\text{diff}}} & \Psi_{24v}^{\text{diff}} & \Psi_{24w}^{\text{diff}} \\[8pt]
    \Psi_{4v1}^{\text{diff}} & \Psi_{4v2}^{\text{diff}} & \Psi_{4v4v}^{\text{diff}} & \Psi_{4v4w}^{\text{diff}} \\[8pt]
    \Psi_{4w1}^{\text{diff}} & \Psi_{4w2}^{\text{diff}} & \Psi_{4w4v}^{\text{diff}} & \Psi_{4w4w}^{\text{diff}} \\
    \end{pmatrix} \, .
\end{equation}
The notation $\Psi_{ij}^{\text{diff}}$ indicates the second derivative of the strain energy computed with the neural network by differentiating the outputs $\hat{\Psi}_i^{\text{p}}$ with respect to the $j^{\text{th}}$ input using back-propagation. We impose positive-definiteness of the Hessian matrix using the principal minor test \cite{prussing1985}. For a matrix to be positive definite, it has to be symmetric and all its leading principal minors, $\Delta_k$, must be positive. This condition is imposed in terms of two additional loss terms,
\begin{align}
    \mathcal{L}_3 &= \frac{1}{N} \sum_{n=1}^{N} \left|\left|\left(\mathbf{H}\right)^{(n)}-\left(\mathbf{H}^\top\right)^{(n)}\right|\right|_F 
    \label{L_sym} \, ,
    \\
    \mathcal{L}_4 &= \frac{1}{N} \sum_{n=1}^{N} \sum_{k=1}^4 \max\left((-\Delta_k^{(n)},0)\right) \, .
    \label{L_LPM}
\end{align}

The total loss is a weighted sum of the terms discussed so far,
\begin{equation}
    \mathcal{L} = \mathcal{L}_1 + a_0\mathcal{L}_2 + a_1\mathcal{L}_3 + a_2\mathcal{L}_4\, .
    \label{Loss_components}
\end{equation}

If training data from sources with different fidelities are used, the total loss of the multi-fidelity (mf) dataset is given as a weighted sum of the losses of the high fidelity (hf) and low fidelity (lf) datasets,
\begin{equation}
    \mathcal{L}_\mathrm{mf} = \mathcal{L}_\mathrm{lf} + a_3\mathcal{L}_\mathrm{hf} \, .
    \label{Loss_mf}
\end{equation}

%The last two terms in (\ref{Loss_components}) can be calculated on a per-datapoint basis without summing over the number of training data points. This is useful because it can then be used as a metric of convexity throughout the input space. Therefore we introduce a new quantity analogous to these, defined as

% \begin{equation}
%     \mathcal{L}_\mathrm{conv}^* = a_1 ||\bar{\mathbf{H}}-\bar{\mathbf{H}}^T||_2 + a_2\sum_{k=1}^4 max(-\Delta_k,0)
%     \label{Loss_star}
% \end{equation}
% where the superscript ($^*$) is used to denote that this quantity is calculated on a point-wise basis.

The training of the neural network was performed using the Adam optimization algorithm \cite{kingma2014adam}. The initial learning rate was set to 0.0001. The exponential decay rates for first and second moment estimates, $\beta_1$ and $\beta_2$, were set to 0.9 and 0.99 respectively. The neural network was trained in 100000 epochs without the use of batching. The training was implemented using Keras \cite{chollet2018keras} with a Tensorflow \cite{abadi2016tensorflow} back-end on a workstation with the following specifications: Intel Xeon E5-1630 3.70 GHz CPU, 16 GB DDR4/2400 MHz random access memory, and Nvidia GeForce GTX 1080 GPU. The values of the weights used in this paper are $a_0=50, a_1=10, a_2=100$ and $a_3=50$. 
%%--------------------------------------------------------------------------------%%
%%% TABLE 1
\begin{table}[h!]\centering
\caption{Neural network architecture}
\label{table01}
\begin{tabularx}{0.52\textwidth}{lll}
%\hline\noalign{\smallskip}
\hline
Layer           & Number of nodes   & Activation function\\ \hline
Input           & 4                 & None\\ 
Hidden layer 1  & 4                 & Sigmoid\\ 
Hidden layer 2  & 8                 & Sigmoid\\
Output          & 5                 & Linear\\
\hline
%\noalign{\smallskip}\hline
\end{tabularx}
\end{table}
%%-----------------------------------------------------------------------%%

%%%%%----- SUBSECTION ---------%%%%
\subsubsection*{Synthetic data generation}
%%%%%--------------------------%%%%
In the majority of biomedical applications it is difficult to obtain sufficient high fidelity data to train a neural network. The number of measurements might be limited, or the data points may be constrained to a narrow region of the input space. It is then beneficial to make use of low fidelity data if available. 
%Here, the high fidelity data is constrained to narrow regions of the input space, such that the performance of the network away from those regions can be very poor. Augmenting the  high fidelity data with easily available low fidelity approximations helps the neural network achieve relatively good performance even in regions with no high fidelity data.

In this study, high fidelity data is in the form of  biaxial stress-stretch measurements. However, only two or three curves within the four-dimensional input space defined by the invariants is explored. Therefore, we generate synthetic data using the Gasser-Ogden-Holzapfel (GOH) \cite{gasser2005GOH} material model. The GOH model proposes an isochoric strain energy of the form
\begin{equation}
    \hat{\Psi} (\mathbf{C}, \mathbf{a}_0) = \hat{\Psi}_{\text{iso}} (\mathbf{C}) + \hat{\Psi}_{\text{aniso}} (\mathbf{C}, \mathbf{a}_0)
\end{equation}
where $\mathbf{a}_0 = (\sin{\theta}, \cos{\theta}, 0)$ is vector denoting the mean fiber direction and parameterized by the angle $\theta$. The functional forms for the GOH strain energy are 
\begin{align}
    &\hat{\Psi}_{\text{iso}}(\mathbf{C}) = \mu (\hat{I}_1-3) \, ,
    \label{GOHpsiiso}
    \\
    &\hat{\Psi}_{\text{aniso}} (\mathbf{C}, \mathbf{a}_0) = \frac{k_1}{4k_2} \left[exp \left(k_2 E^2 \right) -1 \right] \, ,
    \label{GOHpsianiso}
\end{align}

with the generalized fiber strain
\begin{equation}
    E = \left[\kappa \hat{I}_1 + (1-3\kappa) \hat{I}_{4a} - 1 \right]\, .
\end{equation}
The volumetric term that is the same as the one we used to penalize volume changes in our formulation \cite{holzapfel2000, gasser2005GOH, lee2020},
\begin{equation}
    \Psi_{\text{vol}} = \frac{K}{2}(J-1)^2\, .
    \label{GOHpsivol}
\end{equation}

The derivation of the stress tensor for the GOH strain energy is not repeated here, the interested reader is referred to \cite{lee2018, gasser2005GOH}. 
% Following a similar reasoning to Eqs. (\ref{def_S} - \ref{S_to_sigma}) one can show that for a nearly incompressible material the $2^{nd}$ Piola Kirchchoff stress, $\mathbf{S}$, in the GOH material model is given as
% \begin{multline}
%     \mathbf{S} = \mu J^{-2/3} \left ( \mathbf{I} -\frac{1}{3} \hat{I}_1 \mathbf{C}^{-1} \right ) + K(J-1) \mathbf{C}^{-1} 
%     \\
%     + 2k_1 J^{-2/3} E \exp(k_2 E^2) \left[\kappa  \left(\mathbf{I}-\frac{1}{3} \hat{I}_1 \mathbf{C}^{-1}\right) + (1-3\kappa) \left(\mathbf{a}_0 \otimes \mathbf{a}_0 - \frac{1}{3} \hat{I}_4 \mathbf{C}^{-1} \right) \right]
% \end{multline}
% Cauchy stress, $\mathbf{\sigma}$, can then be obtained using (\ref{S_to_sigma}).

Synthetic data with the GOH model is generated by fitting the free parameters to the experimental data using the BFGS optimization algorithm in SciPy \cite{scipy}. 

%%%%%----- SUBSECTION ---------%%%%
\subsubsection*{Finite element method implementation}
%%%%%--------------------------%%%%
We implemented a general neural network material model in a user material subroutine (UMAT) in the nonlinear finite element package Abaqus. The subroutine was written with minimal assumptions to allow for maximal flexibility. The neural network structure, weights and biases, activation functions, etc. are all imported into the subroutine through the input file.

The subroutine performs the following tasks:
\begin{enumerate}
    \item Read in the architecture, weights and biases, activation function types, etc., as a set of material properties.
    \item Pre-process the deformation gradient to obtain the isochoric invariants in Eqs. (\ref{def_iso_invs} - \ref{def_iso_C}).
    \item Perform the forward propagation of the neural network to obtain the predicted strain energy $\Psi^{\mathrm{p}}$ and its first derivatives $\Psi^{\mathrm{p}}_i$.
    \item Calculate stress using Eqs. (\ref{def_S} - \ref{def_Shat}, \ref{S_to_sigma}).
    \item Calculate the second derivatives $\Psi^{\text{diff}}_{ij}$ with back-propagation.
    \item Compute the consistent tangent $\mathbb{c}_{\mathrm{abaqus}}$ using Eq. (\ref{eq_ccabaqus}). 
\end{enumerate}

For the forward propagation,  let $\mathbf{y}_{i-1} \in {\rm I\!R}^{m}$ be the output of layer $i-1$ of the neural network with $m$ nodes. Then the output of layer $i$ is given as

\begin{equation}
    \mathbf{y}_i = g_i(\mathbf{W}_i^T \mathbf{y}_{i-1} + \mathbf{B}_i), \;\;\; \mathbf{y}_i \in {\rm I\!R}^n
\end{equation}
where $g_i:{\rm I\!R} \rightarrow {\rm I\!R}$ is the element-wise activation function, $\mathbf{W}_i \in {\rm I\!R}^{n \times m}$ is the weights matrix, and $\mathbf{B}_i \in {\rm I\!R}^n$ is the biases vector of the $i^{th}$ layer of the network. We use the sigmoid activation function in the hidden layers,
\begin{equation}
    g(y) = \frac{1}{1+e^{-y}}, \;\;\; g'(y) \equiv \frac{dg(y)}{dy} = g(y)(1-g(y))\, .
\end{equation}

For the derivatives, let $\mathbf{J}_{i-1} \in m \times m_0$ be the matrix containing derivatives of the nodes of layer $i-1$ with respect to the inputs of the neural network. Then

\begin{equation}
    \mathbf{J}_i = \text{diag}(g'(\mathbf{y}_i)) \mathbf{W}_i^T \mathbf{J}_{i-1}, \;\;\; \mathbf{J}_i \in {\rm I\!R}^{n \times m_0}
\end{equation}

where $m_0$ is the number of inputs to the neural network and diag$(\bullet)$ denotes a diagonal matrix.

%%%%%----- SUBSECTION ---------%%%%
\subsection*{Biaxial stress-stretch experiments on porcine and murine skin} 
%%%%%--------------------------%%%%

We use experimental data from biaxial stress-stretch experiments performed on murine \cite{meador2020} and porcine skin for the training and validation of neural networks. The data is collected in up to 5 different experimental protocols which are defined in Table \ref{table02}.

%%--------------------------------------------------------------------------------%%
%%% TABLE 1
\begin{table}[h!]\centering
\caption{Experimental loading protocols.}
\label{table02}
\begin{tabularx}{0.3\textwidth}{llll}
%\hline\noalign{\smallskip}
\hline
Loading         & $\lambda_x$       & $\lambda_y$       & $\sigma_z$\\ \hline
Off-x           & $\sqrt{\lambda}$  & $\lambda$         & 0\\ 
Off-y           & $\lambda$         & $\sqrt{\lambda}$  & 0\\ 
Equibiaxial     & $\lambda$         & $\lambda$         & 0\\
Strip-x         & $\lambda$         & 1                 & 0\\
Strip-y         & 1                 & $\lambda$         & 0\\
\hline
%\noalign{\smallskip}\hline
\end{tabularx}
\end{table}
%%-----------------------------------------------------------------------%%

%We use two sets of experimental data in this study. The first set contains the results of biaxial stress-stretch experiments on murine skin \cite{meador2020}. The data in this set were collected under three different loading protocols; equibiaxial, off-biaxial-x and off-biaxial-y. The loading regimes in the protocols are given as $\lambda_x = \lambda_y = \lambda$, $\sigma_z = 0$, in the equibiaxial protocol, $\lambda_y = \lambda, \; \lambda_x = \sqrt{\lambda}$, $\sigma_z = 0$ in the off-biaxial-x protocol (hereafter shortened to off-x), and $\lambda_x = \lambda, \; \lambda_y = \sqrt{\lambda}$, $\sigma_z = 0$ in the off-biaxial-y protocol (hereafter shortened to off-y).

%The second dataset contains the results of off-x and off-y tests on a porcine skin sample from the dorsal region with an equivalent protocol. 

%%-------------------------SECTION-----------------------------%%
\section*{Results}
%%-------------------------------------------------------------%%

%%%%%----- SUBSECTION ---------%%%%
\subsection*{Performance of the neural network against synthetic data}
%%%%%--------------------------%%%%

\begin{figure}
\centering
\includegraphics[width=0.9\linewidth]{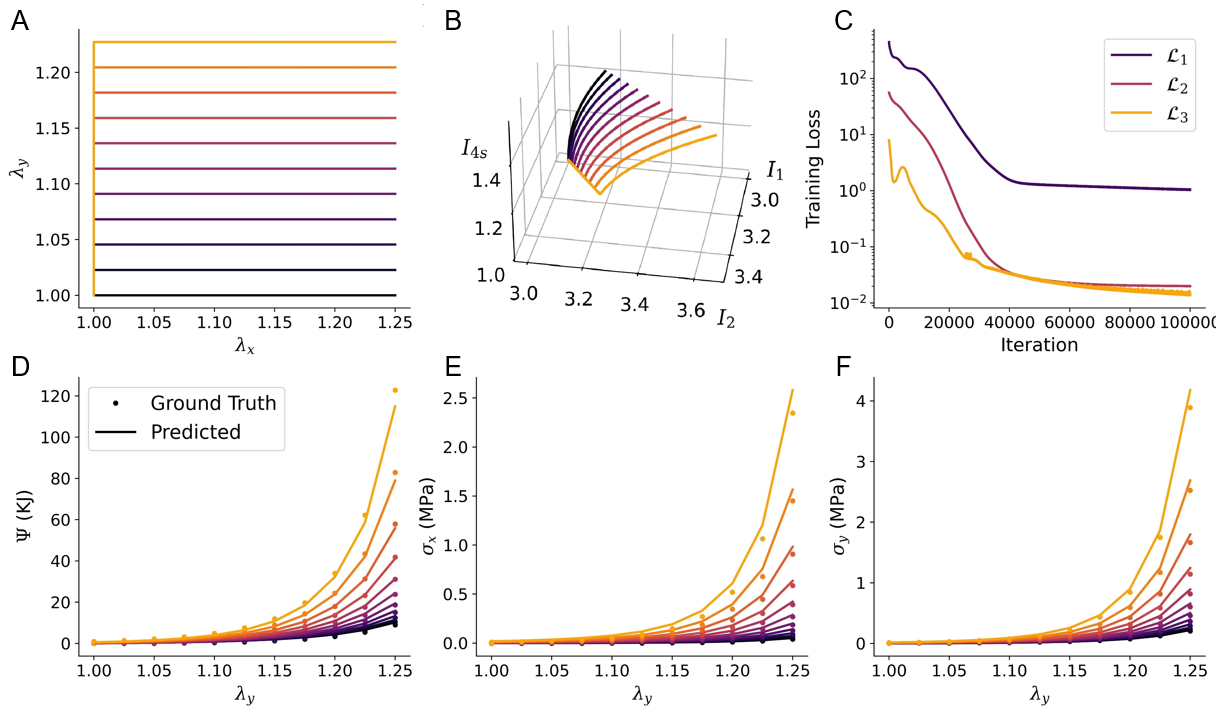}
\caption{Synthetic data generated to train the neural network and the performance of the neural network compared to the training data. (A) Training data was generated by creating \textit{curves} in the $\lambda_x,\lambda_y$  stretch space. (B) The corresponding training data in the invariant space, which is the actual input space for the neural network. The invariant space is four-dimensional but only a three-dimensional projection is shown. (C) Various components of the loss during training. (D) Predicted and ground truth strain energy values throughout the input space. Predicted and ground truth planar stress values in the $x$ (E) and $y$ (F) directions.}
\label{fig_synthetic1} 
\end{figure}

% Paragraph summary: We generate input points (lambdas), transform them to I1,I2..., generate synthetic data, and train the network. The results show that our model is at least as good as GOH
To test the neural network material model, we first train the network using synthetic data only. We generate eleven \textit{curves} in the $\lambda_x , \lambda_y$ space by first holding $\lambda_x = 1$ while $\lambda_y$ is increased gradually to $\lambda_y^{(i)}$, with $i=1,\dots,11$. The values for the $y-$ stretch are $\lambda_y^{(i)}\in[1,1.025,1.05,\dots,1.25]$. After reaching the corresponding $\lambda_y^{(i)}$ value, $\lambda_y$ is held constant while $\lambda_x$ is gradually increased (Figure \ref{fig_synthetic1}A). These loading curves are representative of the the type of test that can be performed experimentally. On the other hand, the neural network takes as inputs the isochoric strain invariants. The invariant space is 4-dimensional, but we plot a 3-dimensional projection in Figure \ref{fig_synthetic1}B. We use the GOH material model to generate synthetic stress data points and train the neural network. Various components of the loss are plotted in Figure \ref{fig_synthetic1}C. The predictions of the trained network are plotted against the training data in Figure \ref{fig_synthetic1}D-F. These results indicate that the neural network is able to recreate expert constitutive models within the training region. 

%%------------------------------------------------------%%
% Fig 2: Performance with synthetic data
%%------------------------------------------------------%%

\begin{figure}
\centering
\includegraphics[width=0.9\linewidth]{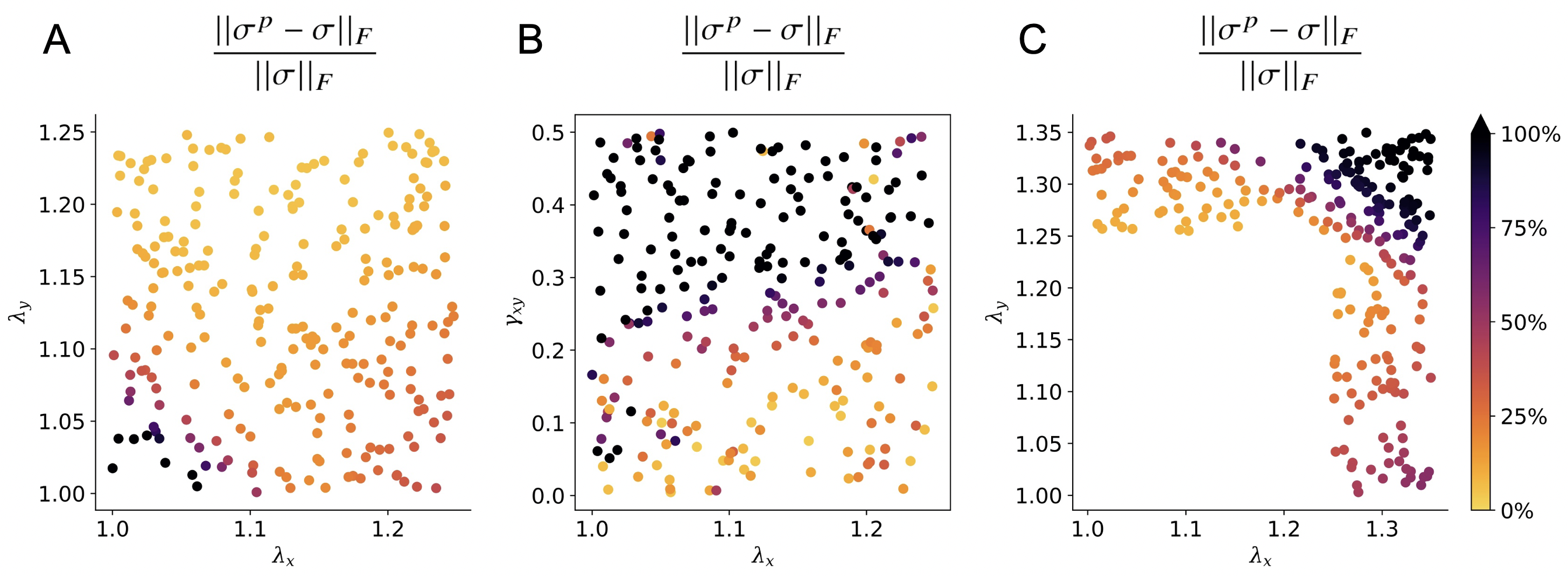}
\caption{Validation of the neural network trained on synthetic data. Performance of the neural network on points randomly sampled from: (A) $\lambda_x\in[1,1.25]$ and $\lambda_y\in[1,1.25]$, (B) $\lambda_x\in[1,1.25]$, $\lambda_y\in[1,1.25]$, $\gamma_{xy}\in[0,0.5]$ and (C) $\lambda_x\in[1,1.25]$ but $\lambda_y\in[1.25,1.35]$; $\lambda_y\in[1,1.25]$ but $\lambda_x\in[1.25,1.35]$; and $\lambda_x\in[1.25,1.35]$ but $\lambda_y\in[1.25,1.35]$ }
\label{fig_GOH_val} 
\end{figure}

% Validation 
We also test if the neural network performs well outside the training region. We generate three validation datasets. The first validation dataset is built by randomly sampling $\lambda_x\in[1,1.25]$, $\lambda_y\in[1,1.25]$ to construct a diagonal deformation gradient of biaxial deformations not seen during training. Then, to test predictions under shear, which are not directly part of the training data, we construct a data set of deformation gradients of the form 

\begin{equation}
    \mathbf{F} = 
    \begin{pmatrix}
    \lambda_x & \gamma_{xy} & 0 
    \\[8pt]
    \gamma_{xy} & \lambda_y & 0
    \\[8pt]
    0 & 0 & \frac{1}{\lambda_x\lambda_y - \gamma_{xy}^2 }
    \\[8pt] 
    \end{pmatrix} .
\end{equation}
The validation dataset is generated from randomly sampling $\lambda_x\in[1,1.25]$, $\lambda_y\in[1,1.25]$, $\gamma_{xy}\in[0,0.5]$. Lastly, we are interested in the potential of the neural network to extrapolate. An additional validation set is constructed by sampling outside the training region: $\lambda_x\in[1,1.25]$ but $\lambda_y\in[1.25,1.35]$; $\lambda_y\in[1,1.25]$ but $\lambda_x\in[1.25,1.35]$; and $\lambda_x\in[1.25,1.35]$ and $\lambda_y\in[1.25,1.35]$. The errors for the validation datasets are shown in Figure \ref{fig_GOH_val}. It can be seen that the neural network performs well within the training region but worse toward the boundary of the training region. Note that the stress values are very small in magnitude, for small deformations (see Figure \ref{fig_synthetic1}D-F); thus, even though the absolute error is small for $\lambda_x,\lambda_y<1.05$, the relative error can achieve close to 100\%. This is not a concern since clearly the stresses at these small deformations are two orders of magnitude smaller compared to stresses at larger deformation. For the third validation set, for example, it can be seen that sampling from $\lambda_x\in[1.25,1.35]$ and $\lambda_y\in[1.25,1.35]$ leads to errors on the order of 300\%. In this region of the input space the stresses are significant. Thus, the neural network has difficulty extrapolating outside of the training region. 

%%------------------------------------------------------%%
% Fig 3: Validation of GOH fit
% $||\sigma^{p}-\sigma||_2 [MPa]$
%%------------------------------------------------------%%

%%%%%----- SUBSECTION ---------%%%%
\subsection*{Performance against experimental data: multi-fidelity data and convexity constraints}
%%%%%--------------------------%%%%

\begin{figure}
\centering
\includegraphics[width=0.98\linewidth]{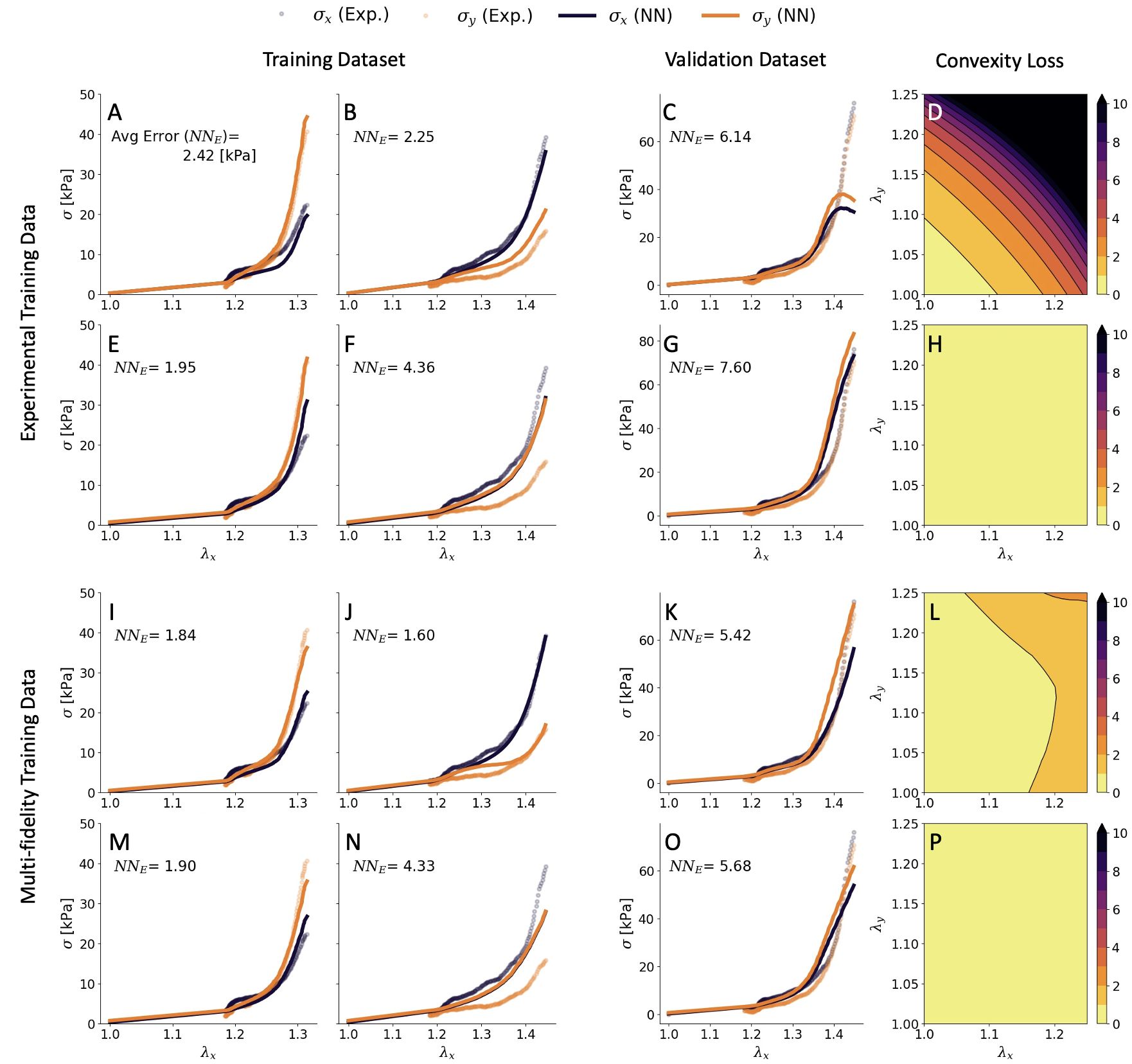}
\caption{Performance of the neural network on the murine skin data. The first two and the third panels of each row show the predicted stress vs actual stress on the training (Off-x and Off-y) and validation (Equibiaxial) sets, respectively, while the fourth panel shows the convexity loss throughout the input space. Each row corresponds to a separate neural network trained differently. Predictions of neural networks trained with (top to bottom): single-fidelity data and no convexity constraints, single-fidelity data and convexity constraints, multi-fidelity data and no convexity constraints, multi-fidelity data and convexity constraints.}
\label{fig_murine} 
\end{figure}

\begin{figure}
\centering
\includegraphics[width=0.95\linewidth]{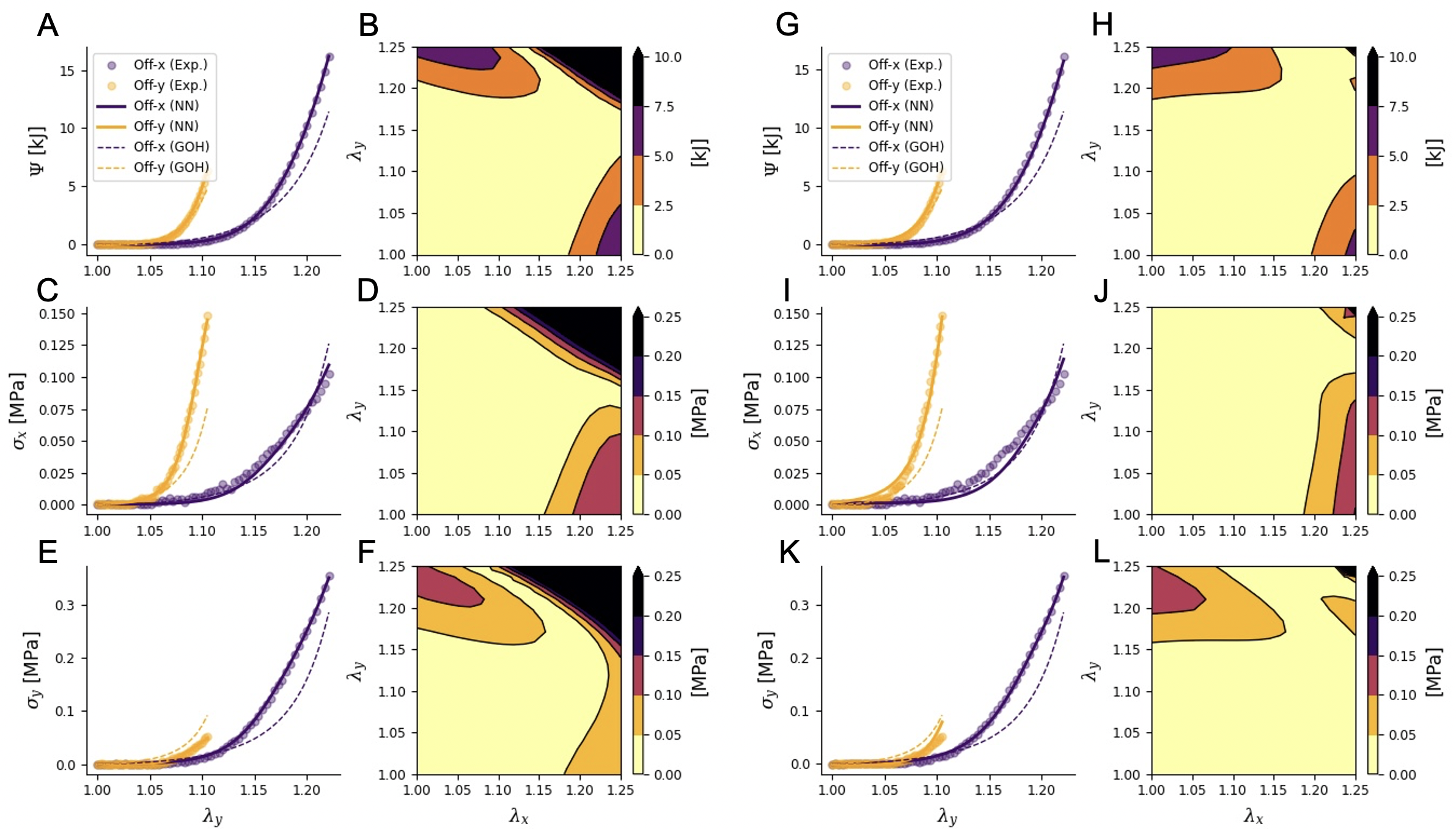}
\caption{Performance of neural networks trained on single-fidelity (A-F) and multi-fidelity (G-L) training data. The scatter plots compare predicted strain energy and stress values to experimental data as well as GOH model outputs. The contour plots show the difference between the corresponding outputs of the GOH model and the neural network.}
\label{fig_porcine} 
\end{figure}

Next, we start training the neural network using experimental data. We want to test the effect of using the experimental data alone (sparse high fidelity data), or combining these data with the low fidelity approximation of the GOH model fit (multi-fidelity data). Concurrently, we want to test if the convexity constraint is required to regularize the fits of the neural network. In Figs. \ref{fig_murine} and \ref{fig_porcine}, we show the results for murine skin data and porcine skin data respectively together with error in the predictions which is defined as the average Frobenius norm of the error in stress, $\mathrm{mean}(||\mathbf{\sigma}^p - \mathbf{\sigma}||_F)$. 

% Re-writing I also think we should: 
% i) calculate the L2 error on the stress for the off x and off y and add it to the figure, also for the biaxial response, so that is is more quantitative 
% ii) make a few changes to the figure (see comment on figure)
The first row of Figure \ref{fig_murine} corresponds to a neural network that is trained using sparse high-fidelity data where the convexity constraints are not imposed. Figure \ref{fig_murine}A shows the neural network ability to fit the off-x and off-y data, achieving average errors of 2.415 kPa and 2.246 kPa, respectively. A biaxial test, not used in training, is used to test the predictive capability of the network (Figure \ref{fig_murine}B). The average error in the validation is 6.137 kPa. Because no convexity constraint is used we can see that it is not satisfied (Figure \ref{fig_murine}C). Keeping only the high-fidelity data but imposing convexity drastically changes the performance. The training loss is poorer (Figure \ref{fig_murine}E and F), but the validation error improves (Figure \ref{fig_murine}G). Moreover, the function is convex over the input space (Figure \ref{fig_murine}H). 

The third and fourth rows of Figure \ref{fig_murine} show the results of neural networks trained with multi-fidelity data. It is notable that even though the network of the third row is trained without any convexity constraints, the fact that it is trained on the GOH synthetic data (which is an inherently convex model) helps it achieve better convexity (Figure \ref{fig_murine}H). Nevertheless, it can be seen that for larger deformations, convexity is lost, as it is evident in the biaxial validation test (Figure \ref{fig_murine}L). The average error in the validation set for the multi-fidelity case without convexity constraint is 5.415 kPa (Figure \ref{fig_murine}K), which is comparable to the sparse high fidelity case without convexity requirements (Figure \ref{fig_murine}C). The results for the neural network trained with multi-fidelity data and with the convexity constraint are shown in last row of Figure \ref{fig_murine}. The training errors are 1.896 kPa and 4.328 kPa (Figure \ref{fig_murine}M and N), and the validation error is 5.678 kPa (Figure \ref{fig_murine}O). Since the convexity is imposed, the loss in the convexity is close to zero over the entire input space (Figure \ref{fig_murine}P). In summary, the convexity constraint is needed, but it can slightly increase the error of the neural network over the training set. This can be reflective of the uncertainties in the experimental data collection, or limitations of the hyperelastic framework. Training with multi-fidelity (given the convexity constraint) yields the best performance: the error in the training data is comparable to the high fidelity case, but the validation error improves. 

%%------------------------------------------------------%%
% Fig 4: Murine data fits
%%------------------------------------------------------%%

%% Table with error, we dont need the GOH comparison here

%As briefly discussed before, augmenting the high fidelity experimental data with low fidelity synthetic data can boost the performance of the neural network. 

%In a multi-fidelity approach, the high fidelity and low fidelity data are weighed differently to enforce that the neural network interpolates the sparse high fidelity data while  using the low fidelity data to guide the neural network in  regions of the input space where no high fidelity data is available. 

In Figure \ref{fig_porcine} we further study the effects of augmenting the training data and how the neural network differs from relying solely on the expert model. For this we focus on porcine skin. We train two neural networks, one of them is trained with the experimental data only (Figure \ref{fig_porcine}A-F), whereas the other is trained on the augmented data (Figure \ref{fig_porcine}G-L). In Figure \ref{fig_porcine}C and D it can be seen that trained only on experimental data, the neural network achieves a low average error of 5.410 kPa, compared to the GOH fit which is 53.164 kPa. Thus, the neural network outperforms the GOH model. This is not surprising since the only task of the neural network is to interpolate the experimental data and satisfy convexity. The contour plots in Figure \ref{fig_porcine}B,D and F shows the difference between the neural network as compared with GOH material model throughout the input space. It can be seen that the two models are very different. Without any data on the rest of the input space, it is unclear if the predictions from the neural network are at all accurate. On the other hand, the GOH model has been developed and trained against thousands of tissue biaxial data. It is reasonable to expect that the GOH model, even though it cannot fit any particular dataset as well as the neural network, it can be trusted to guide the neural network away from the training region. We show that training the neural network on the augmented data, the loss is on average 11.705 kPa against the experimental data (Figure \ref{fig_porcine}C and E), which is still lower with respect to using the GOH model alone. However, as looking at the contours in Figure \ref{fig_porcine}H, J and L we see that the neural network now follows the GOH model closely on the rest of the input space.  

% the previous paragraph is not as important as this paragraph, this is the key, the neural network is the best 'version' of the GOH model, it can be sort of the best hybrid. 
Therefore, the neural network trained with augmented data is at the very least the best version of the GOH model. It performs better than the GOH material model around the high fidelity data points while approximating the GOH model elsewhere. 

%%------------------------------------------------------%%
% Fig 5: Porcine data fits
%%------------------------------------------------------%%

\begin{figure}
\centering
\includegraphics[width=0.98\linewidth]{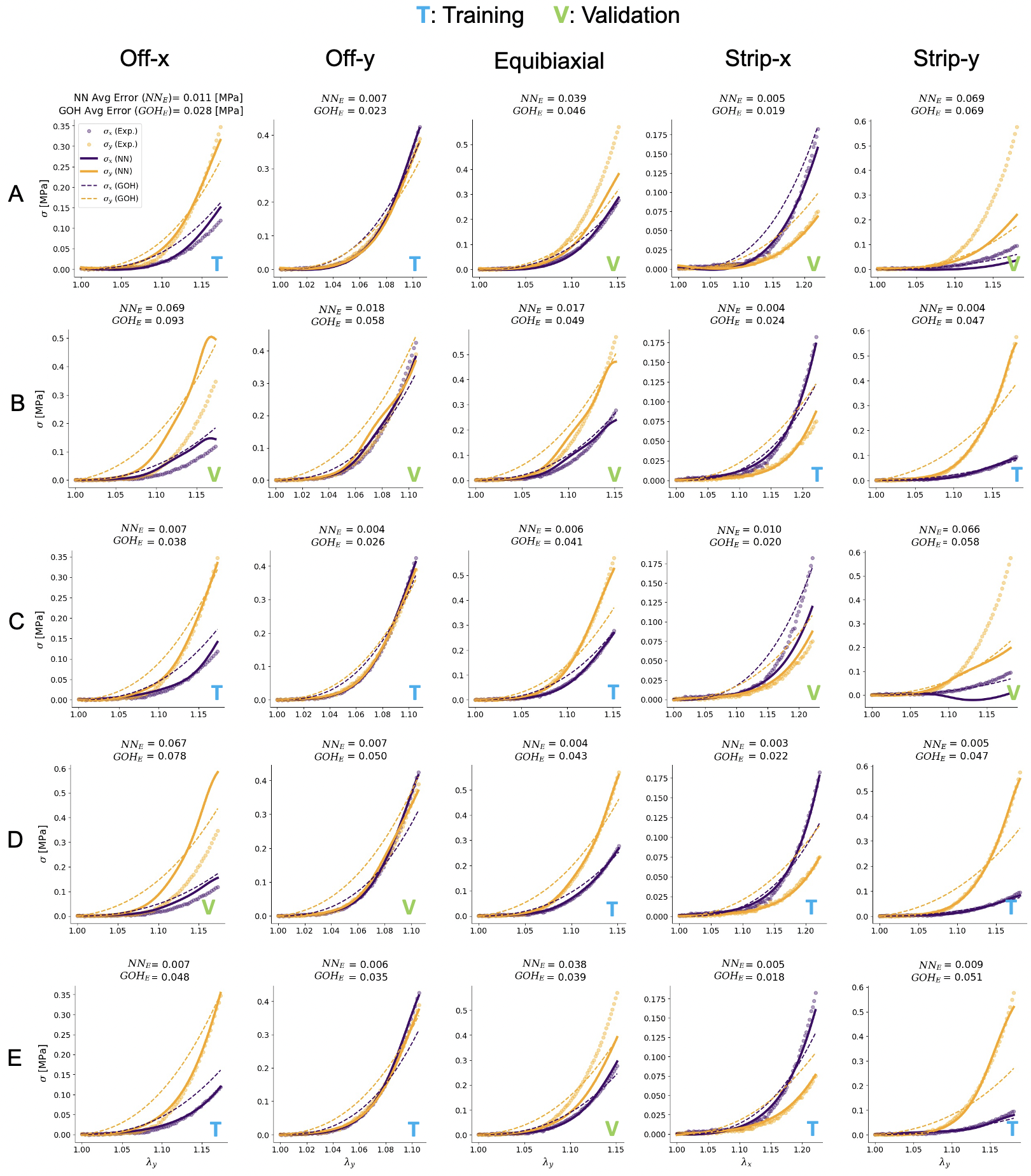}
\caption{Comparison of predicted stress vs actual stress vs GOH fits for various training/validation splits of porcine experimental data. Predictions of a neural network trained on (A) Off-x and Off-y data, (B) Off-x, Off-y and Equibiaxial data, (C) Off-x, Off-y, Strip-x and Strip-y data, (D) Strip-x and Strip-y data, and (E) Equibiaxial, Strip-x and Strip-y data.}
\label{fig_porcineval} 
\end{figure}

The last test of the neural network model is also done with porcine experimental data. In this case we have five different biaxial experiments (see Table \ref{table02}). We are interested in determining which biaxial tests are the most informative for the neural network material model. Thus, we train the neural network with different combination of experimental data and validate against the rest of the data (Figure \ref{fig_porcineval}). We do the same training and testing with the GOH model. In Figure \ref{fig_porcineval}A and B we train the material models with only two datasets, and test against the other three. In the training set, as expected from our previous result, the neural network outperforms the GOH model. Here we see that the neural network continues to outperform the GOH model even in the validation dataset. Figure   \ref{fig_porcineval}C, D show the result of training the models with three of the five biaxial curves, and validating against the remaining two. Again, the neural network obviously outperforms the GOH model in the training set. However, surprisingly, when the neural network is trained with off-biaxial data and tested agains strip-biaxial data, it is now outperformed by the GOH model during the validation step (Figure \ref{fig_porcineval}C). In Figure \ref{fig_porcineval}E we train against four tests, and validate against the equibiaxial test. The validation error is comparable between the neural network and the GOH model. Yet, the superior performance of the neural network on the training data confirms that data-driven models are a preferable alternative to expert constructed constitutive models. 

%%------------------------------------------------------%%
% Fig 6: Porcine data with validation 
%%------------------------------------------------------%%

%%%%%----- SUBSECTION ---------%%%%
\subsection*{Finite element method implementation}
\begin{figure}
\centering
\includegraphics[width=0.98\linewidth]{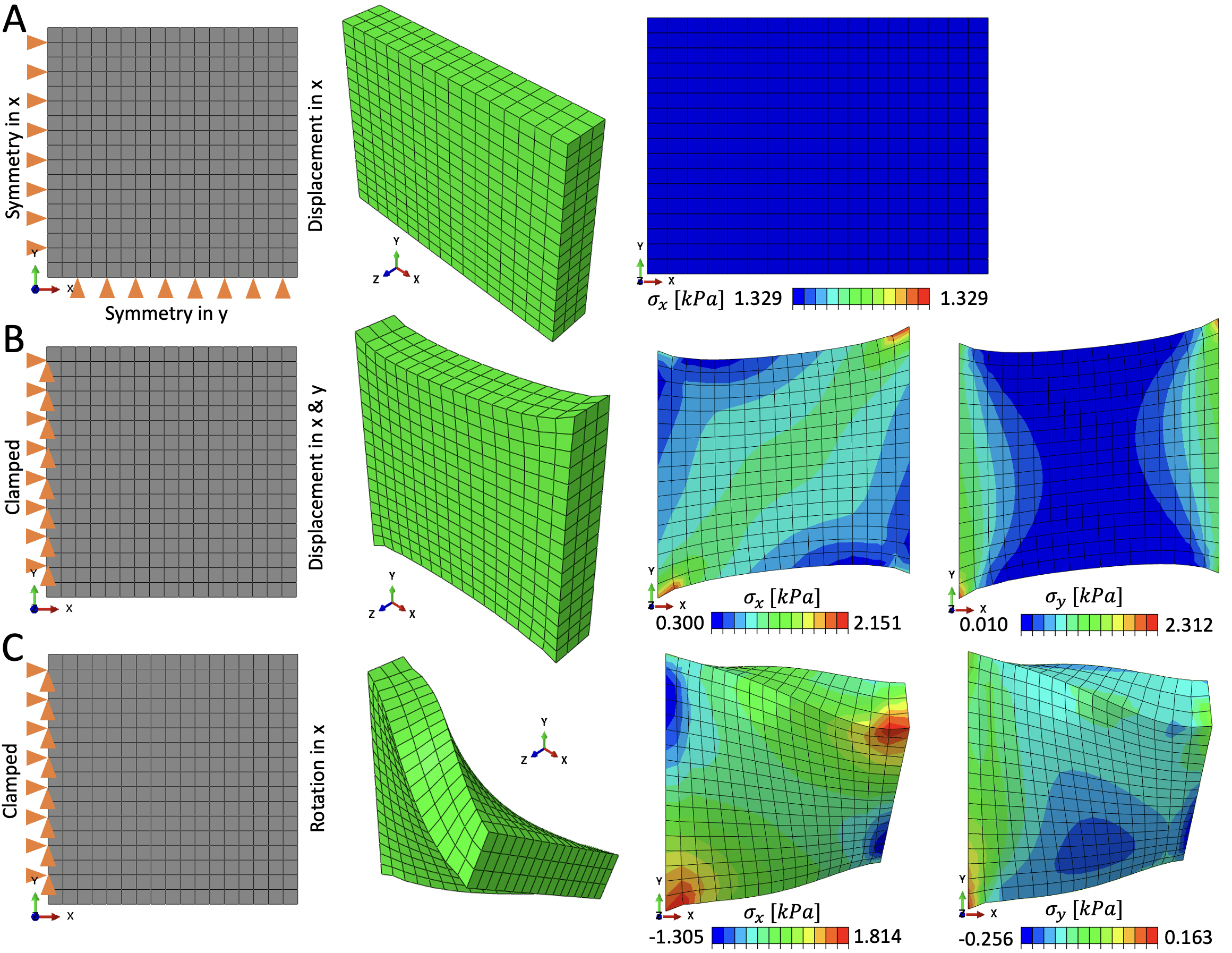}
\caption{Finite element method simulations using the neural network based material model in UMAT. Boundary conditions, deformed geometry and contours of $\sigma_x$ under uniaxial loading (A), Boundary conditions, deformed geometry and contours of $\sigma_x$ and $\sigma_y$ under shear loading (B), and, Boundary conditions, deformed geometry and contours of $\sigma_x$ and $\sigma_y$ under torsional loading (C).}
\label{fig_FEM1} 
\end{figure}

%%%%%--------------------------%%%%
We show a number of basic finite element simulations to test  the capabilities of the neural network material model as a UMAT subroutine for Abaqus. The neural network trained on murine skin data (augmented data with convexity constraint) is defined in the input file. We first consider a rectangular block $5\times5\times1$ cm\textsuperscript{3}. Boundary conditions, mesh and results for a uniaxial extension simulation with $\lambda_x = 1.2$ are depicted in Figure \ref{fig_FEM1}A. The result is a homogeneous stress distribution of $\sigma_x = 1.329$ kPa, $\sigma_y = \sigma_z = 0$, consistent with the results in Fig. \ref{fig_murine},  confirming that the UMAT subroutine is working as intended. 
 
A shearing simulation is shown in Figure \ref{fig_FEM1}B. In this analysis the -x surface of the prism is clamped and a displacement boundary condition of $U_x = U_y = 5$ is applied on the right surface. The contours of the resulting stress components, $\sigma_x$ and $\sigma_y$ are shown in Figure \ref{fig_FEM1}. The Appendix shows a simulation with the GOH fit. As discussed in the previous section, the neural network model with the augmented data is, in a way, the best extension of the GOH model: it retains the expert model features but does not suffer from the constraints of an explicit functional form. 

The last simulation in Figure \ref{fig_FEM1} is a torsional loading scenario. In this simulation the -x surface of the rectangular prism is clamped and a rotation boundary condition of $UR_x = 1 \; rad$ is imposed on the +x surface. The resulting stresses are presented in Figure \ref{fig_FEM1}C. This loading scenario is different from the previous two because it involves significant deformations in the out-of-plane direction. The UMAT subroutine executes without any problems. The three simulations in Figure \ref{fig_FEM1} showcase the robustness and versatility of our neural network UMAT.

\begin{figure}
\centering
\includegraphics[width=0.98\linewidth]{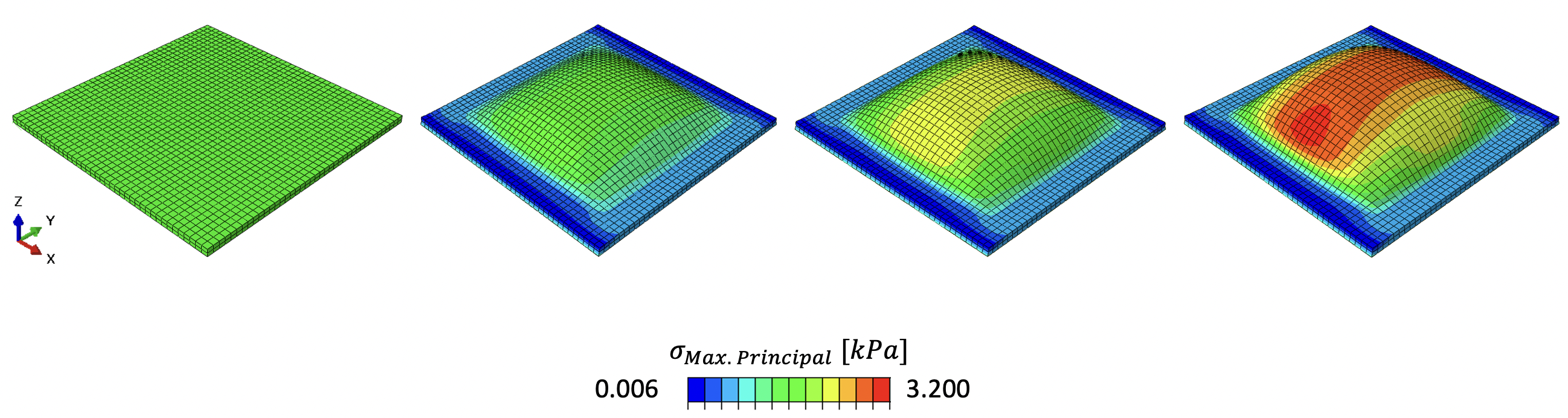}
\caption{Finite element method simulations of tissue expansion using the neural network based material model in UMAT. From left to right: Undeformed geometry, and, contours of maximum principal stress on deformed geometry after the expander is expanded to 20, 40 and 60 cm\textsuperscript{3}, respectively.}
\label{fig_FEM_sqexp} 
\end{figure}

Next, we perform a simulation that is much more closely related to skin biomechanics. Tissue expansion is a widely used technique in reconstructive surgery in which a balloon-like device is inserted and inflated subcutaneously to stretch and grow skin \cite{lee2020}. The domain is a $10 \times 10 \times 0.3$ cm\textsuperscript{3} patch of skin modeled with 3200 brick elements. A rectangular expander of dimensions $8 \times 8$ cm\textsuperscript{2} underneath the skin mesh is modeled with the fluid cavity feature in Abaqus. The expander is inflated to 20, 40 and 60 cm\textsuperscript{3} resulting in the principal strain distributions shown in Figure \ref{fig_FEM_sqexp}. Once again, the simulation converged without issues and the results align with our previous experimental observations of higher deformation at the apex and less toward the periphery of the expander \cite{buganza14}. The simulation in Figure \ref{fig_FEM_sqexp} showcases the ability of our neural network model to be used in realistic finite element simulations through our UMAT. 

\begin{figure}
\centering
\includegraphics[width=0.98\linewidth]{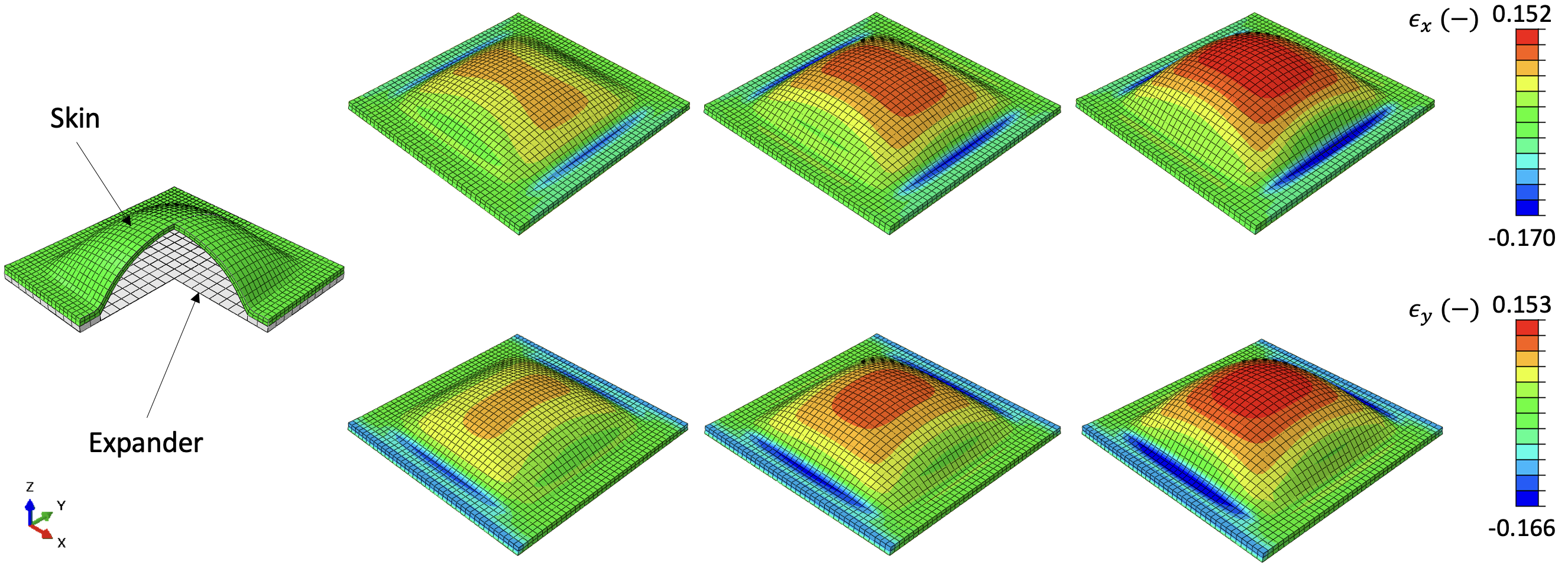}
\caption{Finite element method simulations of tissue expansion using the neural network based material model in the UMAT. From left to right: model setup, and, contours of strain on deformed geometry after the expander is expanded to 20, 40 and 60 cm\textsuperscript{3}, respectively.}
\label{fig_FEM_sqexp2} 
\end{figure}

Note that the simulation in Figure \ref{fig_FEM_sqexp} evidences the anisotropy of the model. The fiber directions $\mathbf{v}$ and $\mathbf{w}$ are aligned with the Cartesian basis $[1,0,0$ and $[0,1,0]$. The tissue is stiffer in the $\mathbf{v}$ direction, which is why there is a band of higher stress along that direction in Figure \ref{fig_FEM_sqexp}. To further showcase the anisotropy in the deformation we also plot the corresponding strains (Figure \ref{fig_FEM_sqexp2}).

%%------------------------------------------------------%%
% Fig 7: FEM 1
%%------------------------------------------------------%%

%%------------------------------------------------------%%
% Fig : FEM 2: expansion
%\begin{figure}
%\centering
%\includegraphics[width=0.98\linewidth]{Figure_7.png}
%\caption{Finite element method simulations of tissue expansion using the neural network based material model in UMAT. From left to right: Undeformed geometry, and, contours of Mises stress on deformed geometry after the expander is expanded to 20, 40 and 60 cm\textsuperscript{3}, respectively.}
%\label{fig_FEM2} 
%\end{figure}
%%------------------------------------------------------%%

%%------------------------------------------------------%%
% Fig 8: FEM 2: square expander
%%------------------------------------------------------%%

%%------------------------------------------------------%%
% Fig 9: square expander
%%------------------------------------------------------%%

\section*{Discussion}
In this study we propose a neural network material model to replace conventional constitutive equations for nonlinear materials, in particular soft collagenous tissues. The neural network takes isochoric strain invariants as inputs and produces the isochoric strain energy and its derivatives with respect to the invariants as outputs. with this design, objectivity is satisfied \textit{a priori}. Other efforts in data-driven modeling of materials and structures rely on training directly on the stress data, which requires additional steps to ensure objectivity \cite{ghaboussi1998, ghaboussi1998_2,kirchdoerfer2016data}. Efforts using invariants or principal stretches have also been shown by others \cite{liu2020, gorieli2018modelselection, dabiri2019}, and by us as well but for isotropic materials \cite{leng2021}. The key ideas introduced in this paper are the consideration of anisotropy, training with multi-fidelity data, convexity constraints, and design of the neural network architecture to compute not only the stress but the consistent tangent needed in finite element simulations. 

Training data for the neural network can consist of both \textit{expensive} (or hard to get) high fidelity data such as results of laboratory experiments,  or a combination of high fidelity data supplemented by synthetic data from expert models. A control case shown here is  to train the neural network on synthetic data alone. The performance of the neural network with the synthetic data shows that the network can interpolate the expert model. This is not surprising since neural networks are universal approximators \cite{leshno1993multilayer}.  When trained on sparse high fidelity data alone, imposing convexity requirements does not only ensure a physically admissible model, but also improves the performance over the validation set (see Figure \ref{fig_murine}). This result reflects the smoothness and convexity of the underlying material behavior. However, even with the convexity regularization, there is no guarantee that the neural network predictions will be accurate outside of the training region. In fact, for the control case in which the neural network is trained on synthetic data alone, predictions outside of the training region are not accurate (see Figure \ref{fig_GOH_val}). 

Unfortunately, high fidelity data of soft tissue mechanics is sparse in most applications. In our previous work, typically only three protocols have been performed: off-biaxial x, off-biaxial y and equi-biaxial (see Table \ref{table02}) \cite{meador2020}. Two other tests, strip-biaxial tests in the x- and y-direction are explored here as well. Liu et al. \cite{liu2020} generated datasets by subjecting tissues to seven biaxial tests. Clearly, more coverage of the input space is always better for data-driven approaches. However, this is a challenge, it requires establishment of multiple repeatable protocols and extensive testing of a individual specimens which can introduce unforeseen uncertainties. Expert models of soft tissue mechanics have evolved over the past few decades and reflect our growing understanding of soft tissues. For example, expert models are often based on microstructure observations \cite{limbert2019skin,kassab2016structure}, are physically meaningful \cite{ehret2007polyconvex,holzapfel2000new}, and have been carefully designed based on observations of many data \cite{Humphrey1990}. On the other hand, expert models can also present limitations, such as non-uniqueness of fit, high sensitivity to parameters, and inability to fit the data due to the inherent constraints of the functional form \cite{tonge2013}. By combining the high fidelity data with an expert model as a low fidelity approximation we aim at getting the best of both, keep data-centric models that can capture the experimental data really well, while at the same time maintaining relatively good performance in regions with scarce high fidelity data.

A strong motivation behind the development of  constitutive models of soft tissues is to be able to build predictive finite element simulations to guide device design or treatment planning \cite{baillargeon2014living}. Previous work on data driven modeling has fallen short in this regard \cite{ghaboussi1998, liu2020, cilla2018}. We implemented the neural network model in a UMAT subroutine for Abaqus, a popular finite element package in both academia and industry. The UMAT subroutine code was implemented with maximum flexibility in mind. The definition and all parameters of the neural network are fed into the UMAT through the input file. We showcased finite element simulations with the neural network trained on the murine data. The simulations converged stably without any issues in all loading scenarios considered, from simple deformations to realistic applications such as tissue expansion. 

Of course, this work is not without limitations. While it is common to model soft tissues within the hyperelastic framework, other physical phenomena will be included in future work, namely viscoelasticity, interstitial flow, and damage. Additionally, a Bayesian framework is needed to account for the inherent uncertainty in material behavior of biological materials. Nevertheless, we anticipate that the general framework introduced here will open up new avenues in data-driven finite element models that balance high-fidelity experimental data with expert knowledge of soft tissue mechanics.  

\section*{Conclusions}
The work presented in this study shows that neural network material models can reliably replace or augment conventional constitutive material models in tissue mechanics analyses. If enough high fidelity data is available, data-driven models can eliminate the burden of choosing a specific functional form and the inherent limitations that come with this choice. However, in most applications, high fidelity data is scarce. Our work demonstrates that a multi-fidelity approach can leverage expert knowledge in the form of synthetic data, while achieving a better fit to the experimental observations. A strong motivation to develop accurate material models of soft tissue is to build predictive finite element models. We designed the neural network with this application in mind, and implemented a neural network UMAT subroutine for Abaqus, a widely used finite element package.

\section*{Data Availability}
Murine biaxial stress-strain data are available through Manuel K. Rausch's dataverse (together with other mechanical raw data): \url{https://dataverse.tdl.org/dataverse/STBML}
Code associated with this manuscript is available at: \url{https://github.com/abuganza/NN_aniso_UMAT}

\section*{Acknowledgements}
This work was supported by the National Institute of Arthritis and Musculoskeletal and Skin Diseases, National Institute of Health, United States under award R01AR074525 to Adrian Buganza Tepole and the National Science Foundation through awards 1916663 and 1916665 to Manuel K. Rausch and Adrian Buganza Tepole, respectively.

% Either type in your references using
% \begin{thebibliography}{}
% \bibitem{}
% Text
% \end{thebibliography}
%

%%%%%%%%%%%%%%%%%%%%%%%%%%%%%%%%%%%%%%%%%%%%%%%%%%%%%%%%%%%%%%%%%%%%%%%%

%% References with bibTeX database:

%\bibliographystyle{model1-num-names}
%\bibliography{References.bib}

%%%%%%%%%%%%%%%%%%%%%%%%%%%%%%%%%%%%%%%%%%%%%%%%%%%%%%%%%%%%%%%%%%%%%%%%%%%%%

\section*{Appendix}

\subsection{Explicit expression for the second Piola-Kirchhoff stress}
\begin{multline*}
    \frac{\partial \hat{\mathbf{S}}}{\partial \hat{\mathbf{C}}} = 
    \delta_1 \mathbf{I} \otimes \mathbf{I} + 
    \delta_2 (\hat{\mathbf{C}} \otimes \mathbf{I} + \mathbf{I} \otimes \hat{\mathbf{C}}) +
    \delta_3 (\mathbf{V}_0 \otimes \mathbf{I} + \mathbf{I} \otimes \mathbf{V}_0 ) + 
    \delta_4 (\mathbf{W}_0 \otimes \mathbf{I} + \mathbf{I} \otimes \mathbf{W}_0 ) + \\
    \delta_5 ( \mathbf{V}_0 \otimes \hat{\mathbf{C}} + \hat{\mathbf{C}} \otimes \mathbf{V}_0) + 
    \delta_6 ( \mathbf{W}_0 \otimes \hat{\mathbf{C}} + \hat{\mathbf{C}} \otimes \mathbf{W}_0) + 
    \delta_7 (\mathbf{V}_0 \otimes \mathbf{W}_0 + \mathbf{W}_0 \otimes \mathbf{V}_0 ) + \\
    \delta_8 \mathbf{V}_0 \otimes \mathbf{W}_0 +
    \delta_9 \mathbf{V}_0 \otimes \mathbf{W}_0 +
    \delta_{10} \hat{\mathbf{C}} \otimes \hat{\mathbf{C}} +
    \delta_{11} \mathbb{I}
\end{multline*}

\begin{align*}
    \delta_1 &= +2(\hat{\Psi}_{11} + 2\hat{I}_1 \hat{\Psi}_{12} + \hat{\Psi}_2 + \hat{I}_1^2 \hat{\Psi}_{22}) \\
    \delta_2 &= -2( \hat{\Psi}_{12}  + \hat{I}_1 \hat{\Psi}_{22} ) \\
    \delta_3 &= +2( \hat{\Psi}_{14v} + \hat{I}_1 \hat{\Psi}_{24v}) \\
    \delta_4 &= +2( \hat{\Psi}_{14w} + \hat{I}_1 \hat{\Psi}_{24w}) \\
    \delta_5 &= -2\hat{\Psi}_{24v} \\
    \delta_6 &= -2\hat{\Psi}_{24w} \\
    \delta_7 &= +2\hat{\Psi}_{4v4w} \\
    \delta_8 &= +2\hat{\Psi}_{4v4v} \\
    \delta_9 &= +2\hat{\Psi}_{4w4w} \\
    \delta_{10} &= +2\hat{\Psi}_{22} \\
    \delta_{11} &= -2\hat{\Psi}_{2}
\end{align*}

\subsection{Finite element method analysis results with GOH material model}

%%------------------------------------------------------%%
% Fig 10: GOH FEM
\begin{figure}
\centering
\includegraphics[width=0.98\linewidth]{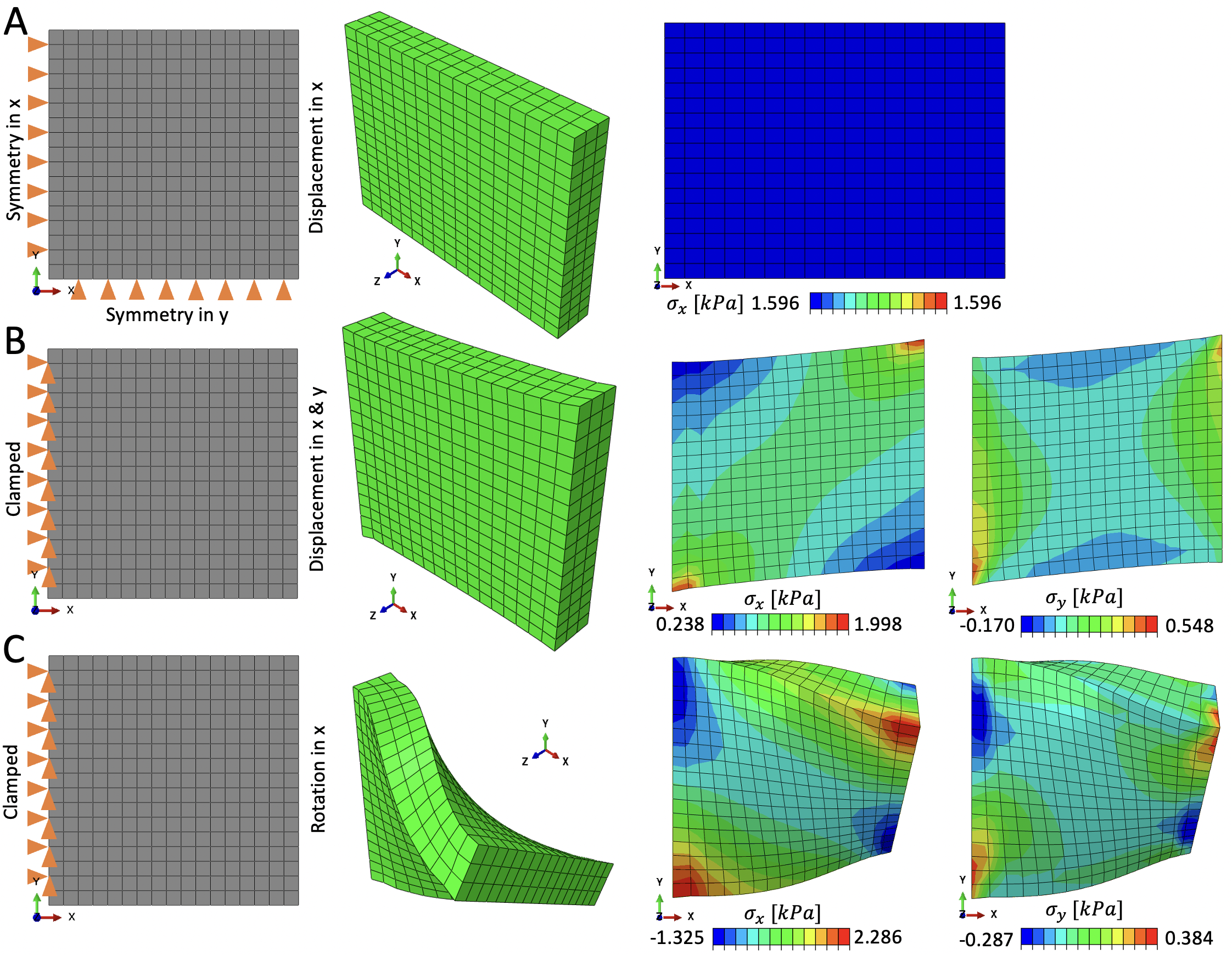}
\caption{Finite element method simulations using the GOH material model in UMAT. Boundary conditions, deformed geometry and contours of $\sigma_x$ under uniaxial loading (A), Boundary conditions, deformed geometry and contours of $\sigma_x$ and $\sigma_y$ under shear loading (B), and, Boundary conditions, deformed geometry and contours of $\sigma_x$ and $\sigma_y$ under torsional loading (C).}
\label{fig_GOHFEM} 
\end{figure}
%%------------------------------------------------------%%
\end{document}